\documentclass[preprint2]{aastex}

\usepackage{txfonts}
\usepackage{graphicx}
\usepackage{amssymb}
\usepackage{epsfig}
\usepackage{natbib}
\usepackage{natbib}
\usepackage{rotating}
\usepackage{prettyref}
\usepackage{textcomp}
\usepackage{color}

\newcommand{\ie}{\emph{i.e.}}
\newcommand{\eg}{\emph{e.g.}}

\newcommand{\SW}{{\it SW}}
\newcommand{\LW}{{\it LW}}
\newcommand{\WideS}{{$Wide$-$S$}}
\newcommand{\WideL}{{$Wide$-$L$}}

\shorttitle{Far-IR imaging of post-AGB and proto-PNe with AKARI/FIS}
\shortauthors{Cox et al.}

\begin{document}
\title{Far-infrared imaging of post-AGB stars and (proto)-planetary nebulae with the AKARI Far-Infrared Surveyor}

\author{N.L.J. Cox}
\affil{Institute of Astronomy, K.U.Leuven, Celestijnenlaan 200D, 3001 Leuven}
\email{nick.cox@ster.kuleuven.be}
\and
\author{D. A. Garc\'ia-Hern\'andez\altaffilmark{1}}
\affil{Instituto de Astrof\'isica de Canarias (IAC), C/ Via L\'actea s/n, E-38200 La Laguna, Tenerife, Spain}
\and
\author{P. Garc\'ia-Lario}
\affil{Herschel Science Centre, European Space Astronomy Centre, ESA, P.O.Box 78, E-28691 Villanueva de la Ca\~nada, Madrid, Spain}
\and
\author{A. Manchado\altaffilmark{1,2}}
\affil{Instituto de Astrof\'isica de Canarias (IAC), C/ Via L\'actea s/n, E-38200 La Laguna, Tenerife, Spain}

\altaffiltext{1}{Departamento de Astrof\'isica, Universidad de La Laguna (ULL), E-38205 La Laguna, Tenerife, Spain}
\altaffiltext{2}{Consejo Superior de Investigaciones Cient\'ificas, Spain}

\begin{abstract}
By tracing the distribution of cool dust in the extended envelopes of post-AGB stars and (proto)-planetary nebulae 
((P)PNe) we aim to recover, or constrain, the mass loss history experienced by these stars in their recent past.
The Far-Infrared Surveyor (FIS) instrument on board the AKARI satellite was used to obtain far-infrared maps 
for a selected sample of post-AGB stars and (P)PNe.
We derived flux densities (aperture photometry) for 13 post-AGB stars and (P)PNe at four far-infrared wavelengths
(60, 90, 140, and 160~$\mu$m).
Radial (azimuthally averaged) profiles are used to investigate the presence of extended emission from cool dust.

No (detached) extended emission is detected for any target in our sample at levels significant with respect to 
background and cirrus emission. Only IRAS\,21046+4739 reveals tentative excess emission between 30 and 130\arcsec.
Estimates of the total dust and gas mass from the obtained maps indicate that 
the envelope masses of these stars should be large in order to be detected with the AKARI FIS.
Imaging with higher sensitivity and higher spatial resolution is needed to detect and resolve, if present, 
any cool compact or extended emission associated with these evolved stars.
\end{abstract}


\section{Introduction}

During the Asymptotic Giant Branch (AGB) phase a low- and intermediate-mass star experiences mass loss (\eg\ by thermal pulses) that
 acts as the main source of replenishment of gas and dust to the interstellar medium
\citep{2005ARA&A..43..435H}. When the AGB star evolves to the post-AGB phase (and later to the proto-planetary nebular (PPN) and
the PN phase) the mass loss drops by several orders of magnitude, while the circumstellar shells continue to drift away
 (pushed by the - episodic - (post)-AGB wind)  from the star into the surrounding medium \citep{2003ARA&A..41..391V}.  
For a constant mass loss rate a smooth envelope with decreasing column density, inverse to the radial distance from the central
star, is expected. If, on the other hand, the mass loss rate fluctuates (as the star goes through thermal pulses
during the AGB;  \citealt{1993ApJ...413..641V}), then enhanced emission should be observed in the form of
multiple  dust shells around the central star. Thus, the history of mass loss during the AGB phase is imprinted
on the dust shell of the PN  until it is disrupted by the fast post-AGB wind. Pre/proto-planetary nebulae
(PPNe) are objects in transition from the high mass loss rate AGB phase to  that of an ionized PN. 
However, little is known yet about the large extended shells around these evolved stars. 
Therefore, young PNe and proto-PNe may have outer shells in which the fossil record of the mass loss during the 
AGB phase is still preserved. 
Cool dust  present in circumstellar shells emits mainly in the infrared. 
Early studies suggested that many PNe are extended in the infrared, but usually with sizes comparable to the optical diameter (\citealt{1991ApJ...374..227H}).
This indicated that the dust grains are generally well mixed with the ionized gas. 
However, IRAS also showed a few PNe with the
far-infrared emission extending beyond the limits derived from optical images. This suggested that cool dust
may be present in a neutral envelope outside the inner ionized zone. The spatial extent and relative brightness
distribution of the mid- and far-infrared radiation of the dust shells around PNe provide important constraints
on the evolution of the central star. In particular the mass loss conditions in the AGB strongly constrain
evolutionary models as this affects the distribution of ionized gas, molecular gas and cool dust in the
resulting PN.

Recent Spitzer/MIPS observations by \citet{2007AJ....134.1419D} have not revealed circumstellar dust shells
around CRL~2688, as reported with ISO. 
The latter were reported by \citet{2000ApJ...545L.145S} who identified dust shells of about 2--3~pc in two post-AGB stars.
\citet{2006A&A...452..523K} tentatively inferred a slightly extended emission from PN
\object{NGC 7008} by means of ISOPHOT multi-aperture photometry. Spitzer/MIPS observations of PN 
\object{NGC 2346} and \object{NGC 650} (at 24, 70 and 160~$\mu$m; \citealt{2004ApJS..154..302S} and
\citealt{2006ApJ...650..228U}, respectively) both show that the 24~$\mu$m morphology is strikingly similar to
that seen in the optical (due to the contribution of nebular emission lines in this band), while the 70 and
160~$\mu$m images show an extended ($\sim$90\arcsec) cold dust envelope around the nebula.
More recently, \citet{2009AJ....138..691C} compared the 24~$\mu$m surface brightness profiles
with those in the H$\alpha$ line for a sample of 36 Galactic PNe;  they found that the infrared emission can be 
more extended, similar or present in the shell center only. 
\citet{2009AJ....138..691C} suggested that these different groups of PNe have an evolutionary connection - with the youngest PNe 
showing  more extended 24$\mu$m emission (due to hot dust)  than the H$\alpha$ emmission.
First results with Herschel reveal detached and extended emission in the far-infrared associated to evolved stars (\citealt{groenewegen2010}) 
as well as wind-ISM interaction at several arcminutes from some objects, such as CW\,Leo (\citealt{2010A&A...518L.141L}).
The latter was also traced in the FUV by GALEX (\citealt{2010ApJ...711L..53S}) similar to the shock interface observed for Mira (\citealt{2007Natur.448..780M}).
\citet{2010A&A...518L.140K} reported detached dust shells  (at $\sim$30-50\arcsec) associated with three AGB stars 
located at 260--825~pc (we note that the most distant AGB star had the dust shell located farthest away from the point source).
Extended - parsec-sized - far-infrared emission is expected to be present outside the ionized region in the neutral zone around
post-AGB stars and PNe as indicated by observations (\citealt{2000ApJ...545L.145S}), as well as theoretical work (\citealt{2002ApJ...571..880V}).
Therefore, if indeed the typical sizes of far-infrared dust emission extends over distances of 2--3~pc, the apparent sizes
would range from  1\arcmin\ to 10\arcmin\ for distances in the range 0.5 to 3~kpc. These are the angular scales we aim to probe 
with the AKARI FIS images. The exact size and shape of the extended emission will also depend strongly not only on the mass-loss rate but also on the relative velocity of the star and the density of the surronding interstellar medium.

In this paper we present a far-infrared survey of thirteen post-AGB stars and (young/proto) PNe obtained with
the Far-Infrared Surveyor (FIS) on the Japanese Infrared Astronomical Mission AKARI.  The properties of the
far-infrared emission of these objects and their immediated surroundings is reported and discussed. 
In Sect.~\ref{sec:observations}
we discuss the observations and the data processing. The main observational results, including photometry, 
are reported in Sect.~\ref{sec:results}.
In Sect.~\ref{sec:discussion} we discuss the main results of this work and the limits on the possible 
presence of extended cold dust emission associated with the post-AGB objects in our sample.
Sect.~\ref{sec:conclusion} concludes the paper.

\section{Infrared mapping with the Far Infrared Surveyor}\label{sec:observations}\label{sec:dataprocessing}

We used the Far Infrared Surveyor (FIS) instrument on board  the AKARI infrared astronomy mission \citep{2007arXiv0708.1796M}
to obtain pointed observations of 13 post-AGB stars and (proto) planetary nebulae (PNe).
The FIS \citep{2004SPIE.5487..359K,2007arXiv0708.3004K} provides scan-maps at four wavelength filter bands with the 
short wavelength (65 and 90~$\mu$m) and long wavelength (140 and 160~$\mu$m) detectors.
The FIS was used in the FIS01 compact source photometry scan-mode. In this mode the detectors sweep the sky twice
in round trips for each pointing. The scan speed was set to 15\arcsec/sec and each scan takes 157.5
seconds to complete. Between two round trips the scan path is shifted by either 70\arcsec\ or 240\arcsec\
resulting in final maps of  10\arcmin\ x 40\arcmin\ or 13\arcmin\ x 40\arcmin, respectively. The sampling
(detector readout) mode was either nominal (with reset interval 0.5 or 1.0 sec) or in CDS mode (which takes 
the ramp differential data: \emph{i.e.} the difference between the begin and end voltage for each interval). The
settings for each pointed observation are given in the observational log (Table~\ref{tb:observation_log}) and
basic target information (ID/names, type, FIS coordinates) is given in Table~\ref{tb:source_properties}.

\begin{table*}
\caption{Properties of the selected sample of post-AGB stars and (proto)-PNe. Observation dates and observing mode settings are also given.}
\label{tb:source_properties}\label{tb:observation_log}
\centering
\resizebox{1\textwidth}{!}{%
\begin{tabular}{lllllllll}\tableline\tableline
IRAS ID		& Alt. Name		 &  Type          & $\alpha$ 	      & $\delta$   	& Date           & FIS       & \multicolumn{2}{c}{FIS01}\\
		&			 &		  &		      &			& dd-mm-year	 & record \# & sampling  & shift (\arcsec) \\
\tableline
07134$+$1005	& \object{HD 56126}	 &  post-AGB/PPN  & 07:16:10.05       & $+$09:59:42.8   & 09-04-2007	 & 4140022    & CDS	  & 240    \\ 
09425$-$6040    & \object{GLMP 260}	 &  post-AGB/C*	  & 09:44:01.63       & $-$60:54:20.2   & 04-01-2007	 & 4140008    & CDS	  & 240    \\ 
10178$-$5958	& \object{Hen 3-401}	 &  post-AGB/PN	  & 10:19:33.12       & $-$60:13:20.0   & 09-01-2007	 & 4140004    & CDS	  & 240    \\ 
		& \object{GLMP 270}	 &		  &		      &			\\
10197$-$5750    & \object{Hen 3-404}  	 &  post-AGB/PPN  & 10:21:31.97       & $-$58:05:36.6   & 07-07-2007	 & 4141023    & 0.5	  & 70     \\ 
		& \object{GLMP 272} 	 &		  &		      &			\\
11478$-$5654    & \object{NGC 3918}    	 &  (CS)PN	  & 11:50:17.11       & $-$57:10:48.1   & 22-07-2007	 & 4141027    & 1.0	  & 70     \\ 
13428$-$6232	& \object{GLMP 363}      &  post-AGB/PPN  & 13:46:20.30       & $-$62:47:56.1   & 15-08-2007	 & 4141020    & 0.5	  & 70     \\ 
15318$-$7144	& \object{Hen 2-131}	 &  (young) PN 	  & 15:37:09.85       & $-$71:54:52.1   & 04-03-2007	 & 4140013    & CDS	  & 240    \\ 
17119$-$5926    & \object{Hen 3-1357}	 &  PN  	  & 17:16:21.00       & $-$59:29:34.9   & 15-09-2006	 & 4140006    & 0.5	  & 240    \\ 
		& \object{GLMP 353} 	 &		  &		      &			\\
17395$-$0841	& \object{GLMP 621}	 &  PN  	  & 17:42:14.11       & $-$08:43:14.5   & 17-03-2007	 & 4140007    & 0.5	  & 240    \\
19500$-$1709	& \object{HD 187885}     &  post-AGB/PPN  & 19:52:53.48	      & $-$17:01:51.5   & 20-10-2006	 & 4140010    & CDS	  & 240    \\ 
20104$+$1950    & \object{NGC 6886}      &  PN  	  & 20:12:43.49       & $+$19:59:32.5   & 03-11-2006	 & 4140015    & 0.5	  & 240    \\ 
		& \object{Hen 2-458} 	 &		  &		      &			\\
21046$+$4739    & \object{NGC 7026}	 &  PN  	  & 21:06:18.87       & $+$47:51:11.3   & 06-06-2007	 & 4141026    & 0.5	  & 70     \\ 
22036$+$5306    & \object{GLMP 1052}	 &  Post-AGB/PPN  & 22:05:30.13       & $+$53:21:31.3   & 28-06-2007	 & 4141021    & 0.5	  & 70     \\ 
\tableline
\end{tabular}
}
\tablenotetext{}{Right Ascension ($\alpha$) and declination ($\delta$) in J2000 coordinates are extracted from the AKARI/FIS \SW\ observations.} 
\end{table*}

\begin{figure}[h!]
\centering
\includegraphics[width=\columnwidth]{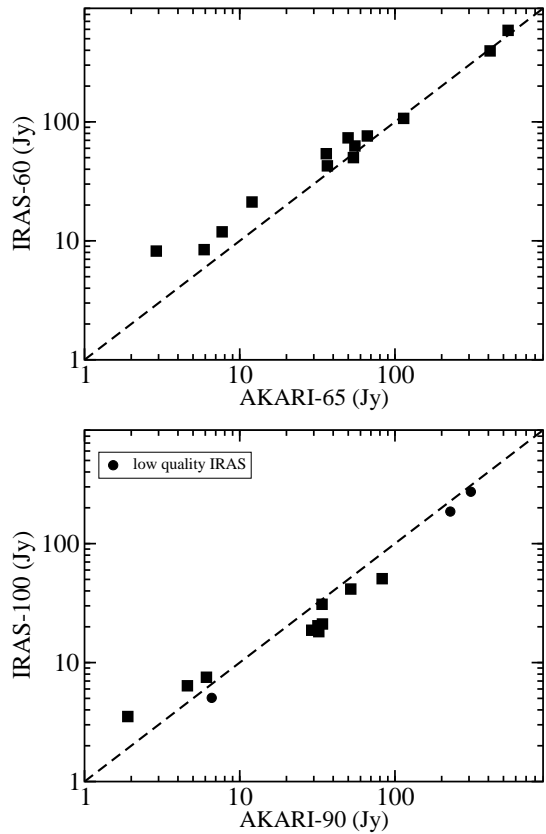}
\caption{%
The flux densities obtained with AKARI (at 65 and 90~$\mu$m; this work) are compared to those given in the
IRAS point source catalog (at 60 and 100~$\mu$m). The dashed line shows the hypothetical (first approximation) one-to-one correlation of 
60 \& 65~$\mu$m and 90 \& 100~$\mu$m fluxes. No correction for color or filter curves has been taken into account. Statistical error bars on 
both AKARI and IRAS point source flux densities are slightly smaller than the symbols.}
\label{fig:compare_flux}
\end{figure}

The pipeline modules provided by the AKARI team (AKARI FIS Data User Manual Version 1.2; \citealt{Verdugo_FIS}) 
were used for the processing of the FIS data. This includes dark subtraction, responsivity correction, flat fielding,
correction for slow transients (\eg\ after calibration lamp exposures).
The pipeline also takes care  of removing gradual long-term changes in detector responsivity,
cosmic-ray glitches and ionizing radiation spikes. 
The distortion of the FIS array detectors field of view and their alignment is also corrected.
The image background level was set to zero by subtracting the sky background in the time domain.
The focal-plane coordinate system was converted into the ecliptic coordinate system.
In this way all frames have the same orientation since the scanning is done parallel to the ecliptic plane.
The images are created with grid sizes of 30\arcsec\ for the \WideL\ and $N160$ bands
and 15\arcsec\ for the \WideS\ and $N60$ bands.
We verified that the positions of the observed targets (obtained via a 2D Gaussian fit to the infrared point source) 
agree within $\sim$10\arcsec\ with those of the optical/near-infrared 
counterparts (\eg\ 2MASS/Simbad: \citealt{2003yCat.2246....0C}, \citealt{2003yCat..34081029K}).

\begin{table*}
\caption{FIS filter \& detector properties.}\label{tb:filter_detector}\label{tb:correctionfactors}
\centering
\begin{tabular}{lllll}\tableline\tableline
Band	 		 & $N60$    	& \WideS\	& \WideL\	& $N160$	\\\tableline
$\lambda$ ($\mu$m) 	 & 65		& 90	 	& 140		& 160		\\
$\Delta\lambda$ ($\mu$m) & 22		& 38		& 52		& 34		\\
wavelength range ($\mu$m)& 50-80	& 60-110	& 110-180	& 140-180	\\
FWHM ('')		 & 32$\pm$1	& 30$\pm$1	& 41$\pm$1	& 38$\pm$1	\\
Pixel size ('')		 & 26.8		& 26.8		& 44.2		& 44.2		\\
\tableline
\smallskip
diffuse-to-point	& 0.698 (flux)$^{-0.0659}$ & 0.700 (flux)$^{-0.0757}$  & 0.560 	& 0.277  	\\
aperture correction	& 0.882		& 0.865   	& 0.881 	& 0.910 	\\ 
\tableline
\end{tabular}
\tablenotetext{}{See \citet{Verdugo_FIS} for additional details.
Also listed are the adopted correction factors to obtain absolute flux levels. 
Aperture correction values are given for aperture radii of $\sim$75\arcsec\ and $\sim$90\arcsec\ for the \SW\ and \LW, respectively
(from \citealt{2009PASJ...61..737S}).}
\end{table*}

\begin{figure*}
\centering
\includegraphics[bb=30 140 300 800,width=8cm,height=13.5cm,clip]{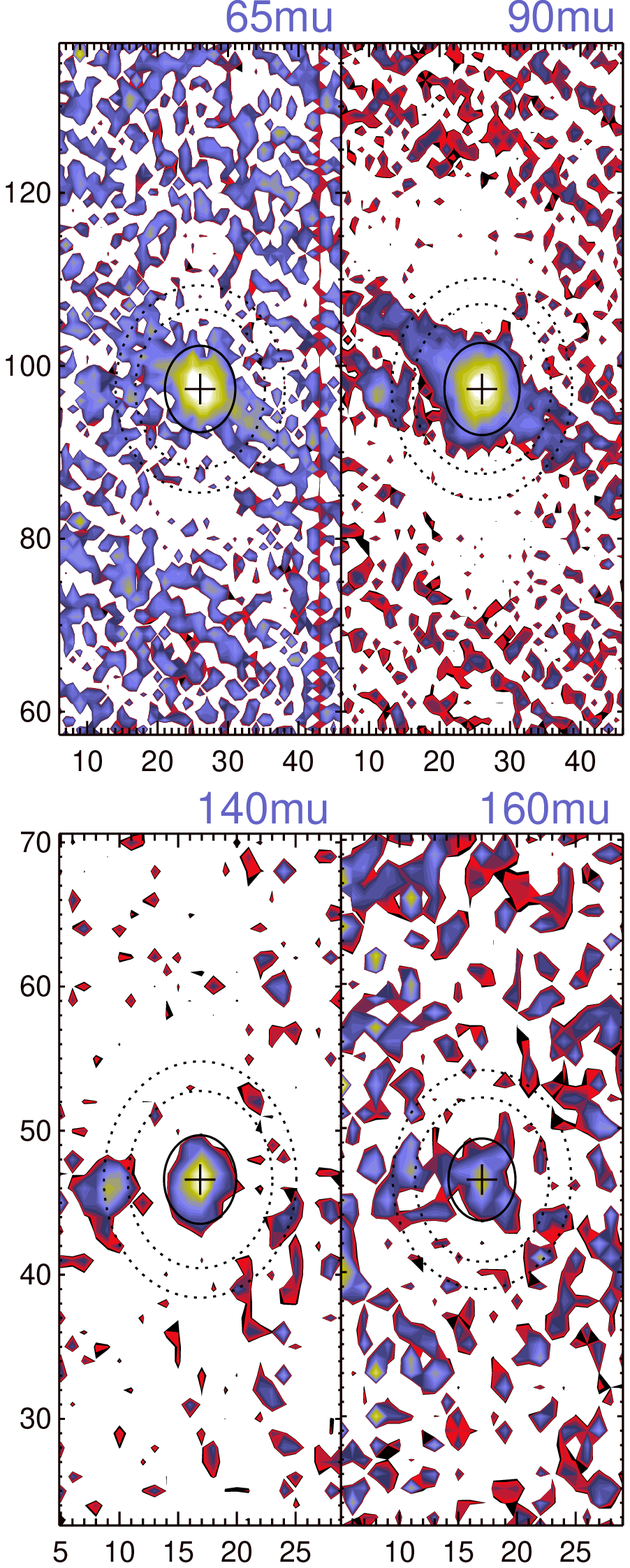}
\includegraphics[bb=30 140 300 800,width=8cm,height=13.5cm,clip]{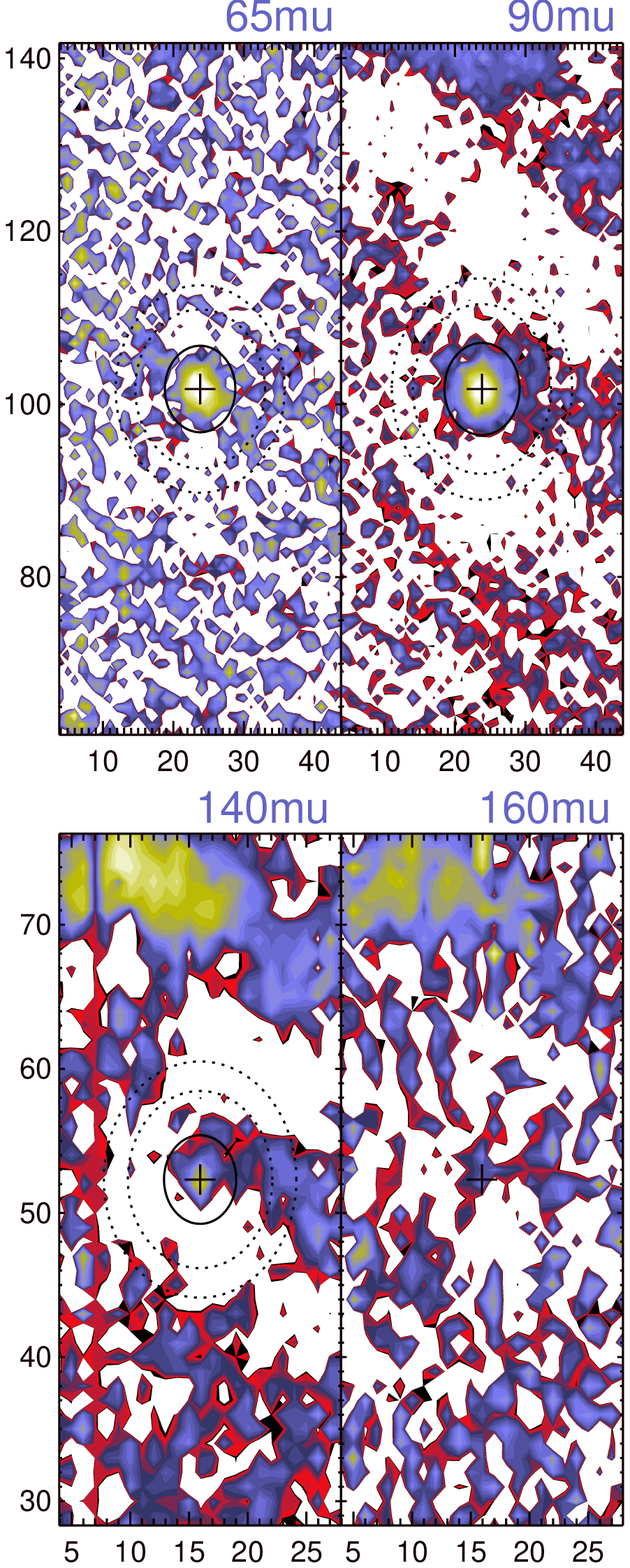}
\caption{The final images in the 60, 90, 140 and 160~$\mu$m band for IRAS\,07134+1005 and for IRAS\,09425-6040.
Contour levels are drawn in logarithmic intervals (scaled between minimum and maximum brightness).
The crosses show the source central position. 
All images are oriented with the scan direction upwards (\ie\ along the ecliptic) which facilitates comparisons.
Pixel sizes are 15\arcsec\ and 30\arcsec\ for the \SW\ and \LW\ bands, respectively (Note that the axes labels are in pixel units).
Also shown are the aperture and sky annuli used for the aperture flux photometry.
Note that the strong bar (angle of $\sim$50\degr\ with respect to the ecliptic) feature in the \SW\ band is an detector artifact (see
\eg\ \citet{2009PASJ...61..737S}. It is not present in the \LW\ images.
For IRAS\,09425-6040 the 140 and 160~$\mu$m images reveal a cold dust emission feature, with peak temperature around 20~K, located 10~\arcmin\ 
upwards (in the direction parallel to the ecliptic) of the central source. Potentially, this is an old detached shell, 
but we cannot rule out diffuse cirrus emission.}
\label{fig:fismap}
\end{figure*}

The nominal FWHM of the PSF is $\sim$40\arcsec\ for the \LW\ and $\sim$30\arcsec\ for the \SW\ detectors, respectively. 
Relevant filter and detector properties are given in Table~\ref{tb:filter_detector}. 
More details on the FIS and its performance are given in \citet{2004SPIE.5487..359K,2007PASJ...59S.389K} and \citet{2009PASJ...61..737S}.
In effect, for the \LW\ we applied the transitory correction and applied a smooth filter (90s boxcar) with 1.5\,$\sigma$-clipping.
For the \SW\ we applied in addition (for the non-CDS observations) a median filter (200s) and used the local flat.
The reduced images for all targets are shown in Figure~\ref{fig:fismap}.

\begin{table*}[ht!]
\caption{Photometric fluxes $F(\lambda)$ and statistical errors $\sigma_F$ in Jy for each target 
for each of the AKARI FIS filters.}
\label{tb:photometry}
\centering
\resizebox{1\textwidth}{!}{%
\begin{tabular}{lllllllllrr}\tableline\tableline
	       & \multicolumn{8}{c}{AKARI FIS Flux Density (Jy)} & \multicolumn{2}{c}{3$\sigma$ detection limit}\\ \tableline
IRAS name      & F(65~$\mu$m) & $\sigma_F$ & F(90~$\mu$m) & $\sigma_F$ & F(140~$\mu$m) & $\sigma_F$ &  F(160~$\mu$m) & $\sigma_F$ & 90~$\mu$m & 140~$\mu$m \\ \tableline
07134$+$1005   &   54.0   &  1.3   & 29.1   &  0.3   &  5.9    & 0.5   & 5.0	&  1.5     & 0.4	 & 0.3	       \\
09425$-$6040   &   12.0   &  0.9   & 6.6    &  0.2   &  2.1    & 0.4   & --	&	   & 0.4	 & 0.4         \\
10178$-$5958   &   66.4   &  1.4   & 51.9   &  0.6   & (18.5)  & 2.7   &(13.9)  &  5.5     & 0.7	 & 1.3	       \\
10197$-$5750   &  535.6   &  2.5   &308.4   &  2.7   & (48.6)  & 9.3   & --	&	   & 5.1 	 & 14.1        \\
11478$-$5654   &   36.1   &  0.1   & 31.9   &  0.1   &  6.7    & 0.4   & 5.6	&  0.9     & 0.2	 & 0.2	       \\
13428$-$6232   &  410.1   &  3.1   & 227.3  &  2.4   &  --     &       & --	&	   & 5.4 	 & 17.6        \\
15318$-$7144   &   55.2   &  1.4   & 34.1   &  0.3   &  10.2   & 0.4   & 7.9	&  2.7     & 0.6	 & 0.3	       \\
17119$-$5926   &   2.9    &  0.1   & 1.9    &  0.1   &  0.8    & 0.2   & --	&	   & 0.06	 & 0.2	       \\
17395$-$0841   &   5.9    &  0.1   & 4.6    &  0.1   & (1.3)   & 0.5   & --	&	   & 0.1	 & 0.4	       \\
19500$-$1709   &   49.9   &  0.9   & 32.3   &  0.2   &  7.3    & 0.2   & 4.0	&  0.6     & 0.5	 & 0.3	       \\
20104$+$1950   &   7.7    &  0.1   & 6.1    &  0.1   &  2.8    & 0.3   &(2.5)	&  0.5     & 0.08	 & 0.3	       \\
21046$+$4739   &   36.7   &  0.1   & 33.9   &  0.1   &  15.5   & 0.6   & 12.1	&  0.9     & 0.2	 & 0.8	       \\
22036$+$5306   &  113.8   &  0.2   & 82.7   &  0.1   &  24.9   & 0.3   & 19.2	&  0.7     & 0.2	 & 0.3         \\
\tableline
\end{tabular}
}
\tablenotetext{}{Values in parenthesis are less secure due to possible low-level `cirrus' contamination.
The 3$\sigma$ detection limits on the extended flux are in Jy/arcmin$^2$.}
\end{table*}

\section{Results and data analysis}\label{sec:results}

From the reduced images we extract information on the total point source flux density and detection limits on the extended emission.
Azimuthally averaged radial profiles were constructed in order to search for extended spherical structures associated with faint extended dust emission.

\subsection{Absolute point source aperture photometry and detection limits on the extended emission}\label{subsec:photometry+sed}
Circular aperture photometry was applied to all sources to derive absolute fluxes.
For the \SW\ fluxes the aperture radius was set to 5 pixels (or $\sim$75\arcsec). 
The sky background was determined from an annulus with inner and outer radii of 9 and 12 pixels, respectively.
For the \LW\ flux measurements we adopted an aperture radius of 3 pixels ($\sim$90\arcsec).
The sky background was computed from an annulus with inner and outer radii of 6 and 8 pixels, respectively.
To convert the obtained image fluxes to absolute fluxes we apply the diffuse-to-point correction factor as well as the 
aperture correction factor (see Table~\ref{tb:correctionfactors} and \citealt{2009PASJ...61..737S}).
Note that the flux calibration by aperture photometry refers to a flat ($\nu F_\nu$ = constant)
spectrum while the real flux depends on the spectral energy distribution (SED) of the source. 
At this stage we have not applied any color correction.
\citet{2009PASJ...61..737S} report an absolute FIS calibration accuracy better than 15\%, except for the longest wavelength ($\sim$50\%). 
These authors also report a tentative 10\% decrease in observed flux in all bands for the fast (15\arcsec/s) scan speed
with respect to the nominal 8\arcsec/s scan speed. 
As all our targets were obtained with 15\arcsec/s this may result in a small underestimation of the final flux densities. 
Furthermore, we assume these calibration factors are valid also for  CDS mode (recommended for very bright sources) observations although this has not been verified in depth. 
Our results can be compared directly with the spectral energy distribution provided for 5 targets included in the 
{\it Toru\'n catalogue of Galactic post-AGB and related objects} (\citealt{2007A&A...469..799S}). 
From these SEDs there  appears to be no evidence for systematic lower (or higher)  than expected flux densities.
Figure~\ref{fig:compare_flux} illustrates that the absolute fluxes for bright point sources measured with AKARI FIS at 65 and 90~$\mu$m are, at a first approximation,
consistent with the IRAS fluxes at 60 and 100~$\mu$m, even without applying a color correction.
For the post-AGB/proto-PNe in our sample, nebular lines are expected to contribute little to the FIR broad band emission;
this effect could be more significant for the PNe.
This consistency is also an important calibration check for flux densities reported in the AKARI bright source catalog (Yamamura et al. 2009).

For the 90 and 140~$\mu$m maps we also measure  the average flux noise level (Jy arcmin$^{-2}$) from the sky background within 2 to 5\arcmin\ from
the point source. The 3$\sigma$ detection limits derived from our FIS maps for each target are listed in Table~\ref{tb:photometry}.
These upper limits are used in Sect.~\ref{sec:discussion} to derive upper limits on the dust and gas mass in the extended envelopes.

\begin{figure*}[ht!]
\centering
\includegraphics[width=1.75\columnwidth,clip]{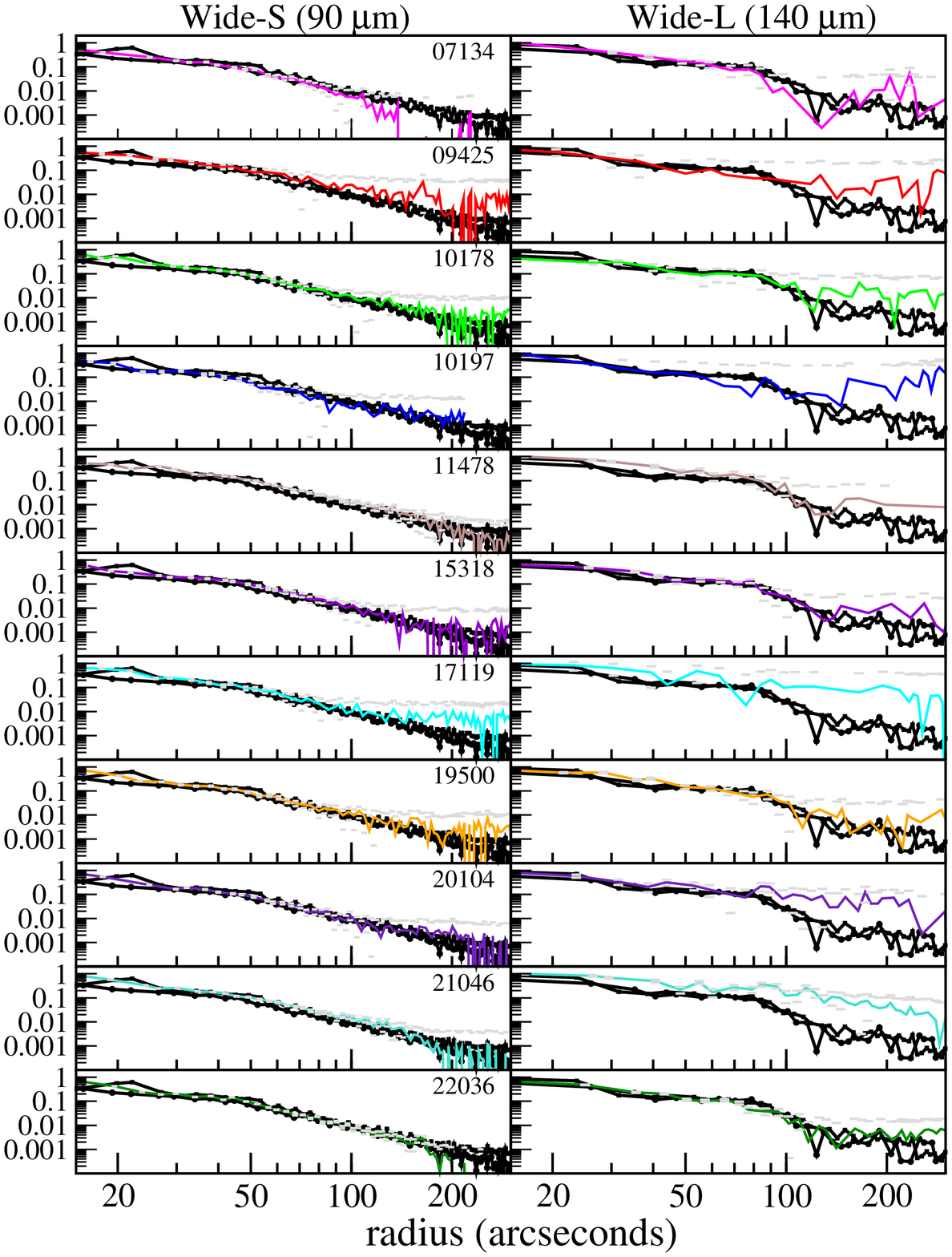}
\caption{%
The azimuthally averaged normalized intensity radial profile for the point source calibration sources 
(Neptune and Vesta in black) and for our sample of post-AGB and (P)PNe sources for the Wide-S and Wide-L bands.
Wide-S and Wide-L profiles shown in the left and rigth column, respectively.
Targets are indicated with their first 5 IRAS digits in each row. Error bars are indicated by 
grey horizontal lines above/below the coloured curve. Due to logarithmic scaling negative values are not displayed.
Thus if no error bars are visible below the curve this implies that they fall below the y-axis ordinate of 0.0001.
Note that only for IRAS\,21046$+$4739 the uncertainties are small enough to hint at extended excess emission at radii 
between $\sim$30--130\arcsec.}
\label{fig:radprofiles}
\end{figure*}

\subsection{Azimuthally averaged radial intensity profiles}\label{subsec:images}

We extracted the azimuthally averaged radial profile for each target in each filter band. 
In addition, we extract radial profiles from calibration point sources (Vesta and Neptune) processed  in the same manner as our targets.
The radial profiles for the calibration point sources and all post-AGB and PNe sources are shown on a log--log scale in
Figure~\ref{fig:radprofiles} for the two most senstive bands at 90 and 140~$\mu$m.
Due to the width of the image maps proper radial profiles can only be obtained out to radii of about 300\arcsec.
As indicated above, there is no unambiguous evidence for extended emission within 2\arcmin\ of the point source.
The sensitivity of these maps is insufficient to unambiguously detect weak detached emission between 2\arcmin~and~5\arcmin.
One potential exception is IRAS\,21046+4739, for which localized far-infrared emission at 140~$\mu$m (and also present in the 160~$\mu$m map) 
could be related directly to the PN.

\clearpage

\clearpage

\addtocounter{figure}{-1}

\begin{figure*}[t!]
\centering
\includegraphics[bb=30 140 300 800,width=5cm,height=8.6cm,clip]{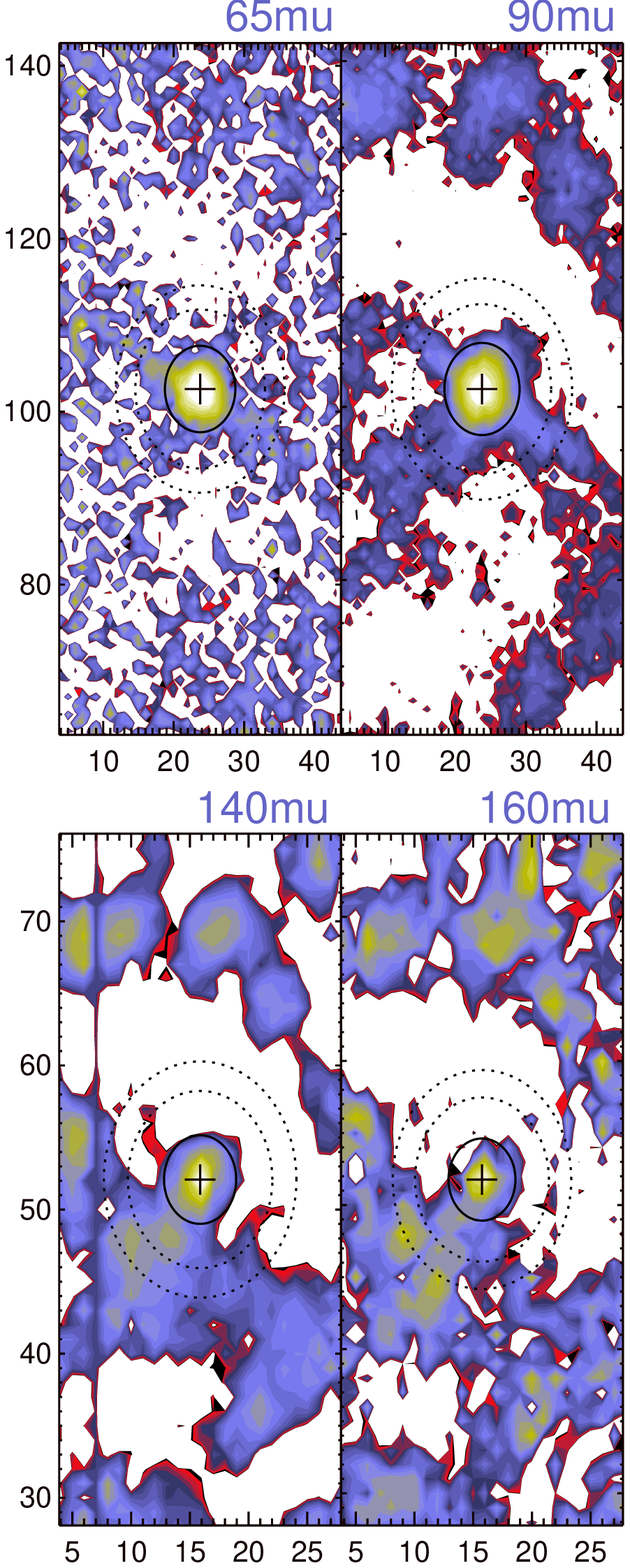}
\includegraphics[bb=30 140 300 800,width=5cm,height=8.6cm,clip]{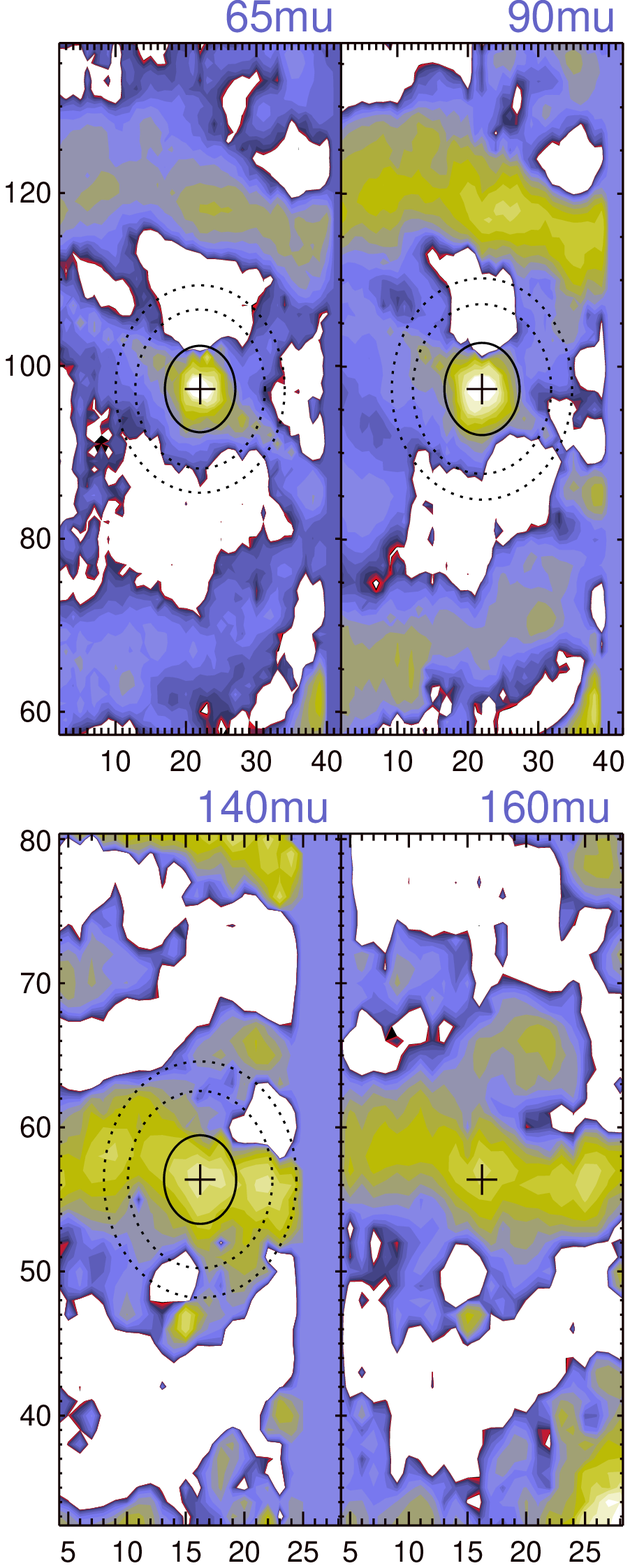}
\caption{%
(continued). For IRAS\,10178-5958 and IRAS\,10197-5750.}
\end{figure*}
\addtocounter{figure}{-1}

\begin{figure*}[t!]
\centering
\includegraphics[bb=30 140 300 800,width=5cm,height=8.6cm,clip]{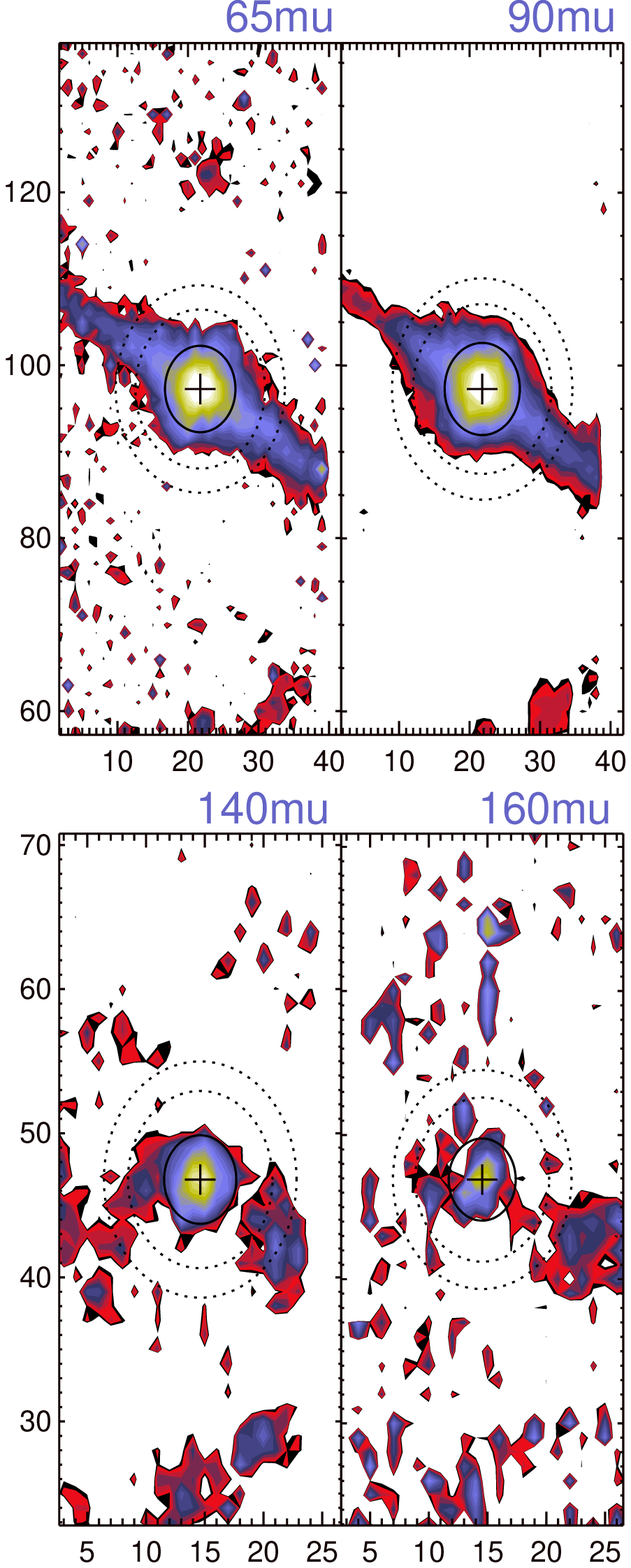}
\includegraphics[bb=30 140 300 800,width=5cm,height=8.6cm,clip]{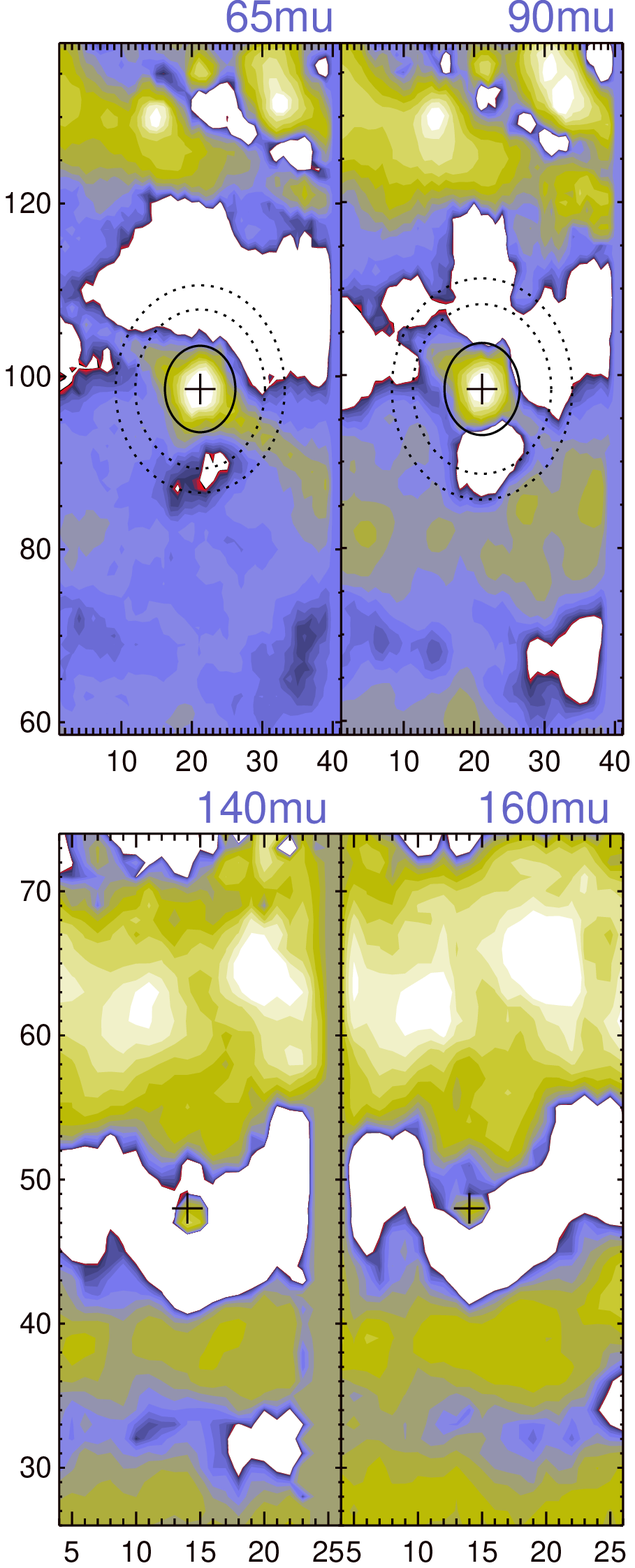}
\label{fig:images}
\caption{%
(continued). For IRAS\,11478-5654 IRAS\,13428-6232.}
\end{figure*}
\clearpage
\addtocounter{figure}{-1}

\begin{figure*}
\centering
\includegraphics[bb=30 140 300 800,width=5cm,height=8.6cm,clip]{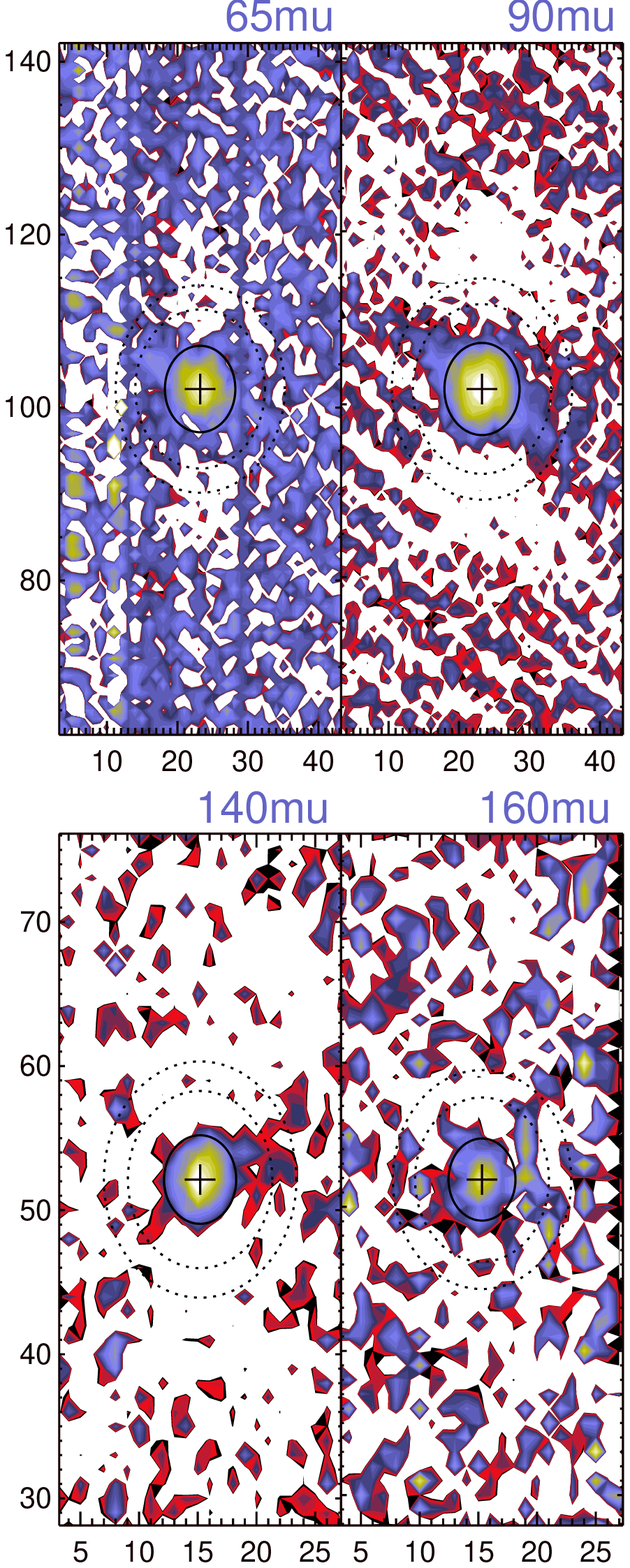}
\includegraphics[bb=30 140 300 800,width=5cm,height=8.6cm,clip]{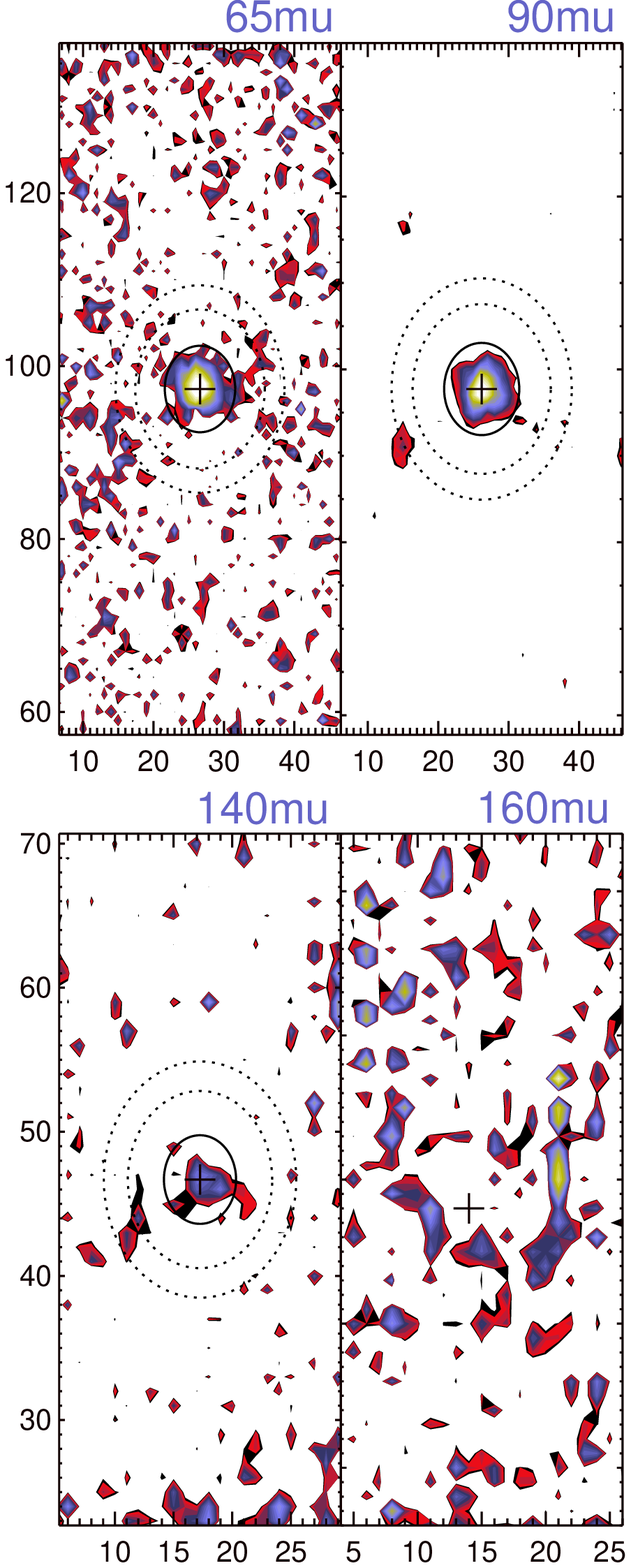}
\caption{(continued). For IRAS\,15318-7144 and IRAS\,17119-5926.}
\end{figure*}
\addtocounter{figure}{-1}

\begin{figure*}
\centering
\includegraphics[bb=30 140 300 800,width=5cm,height=8.6cm,clip]{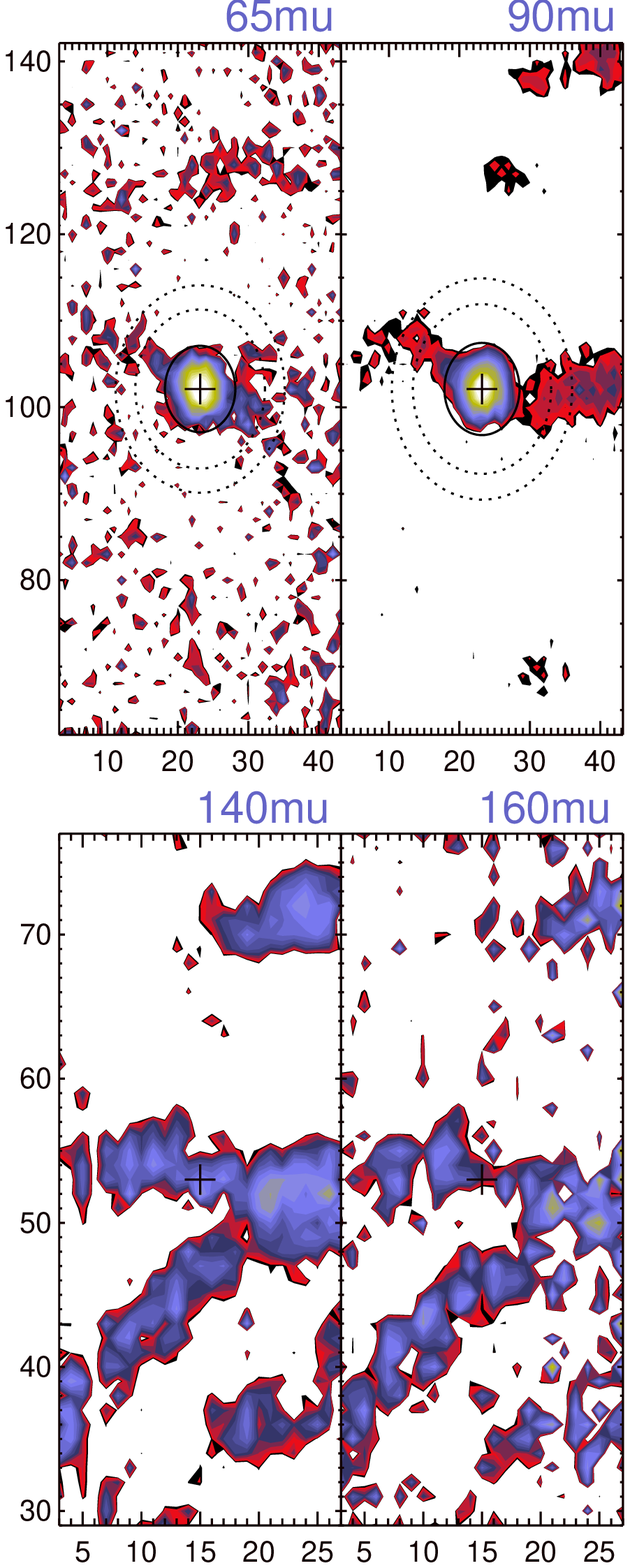}
\includegraphics[bb=30 140 300 800,width=5cm,height=8.6cm,clip]{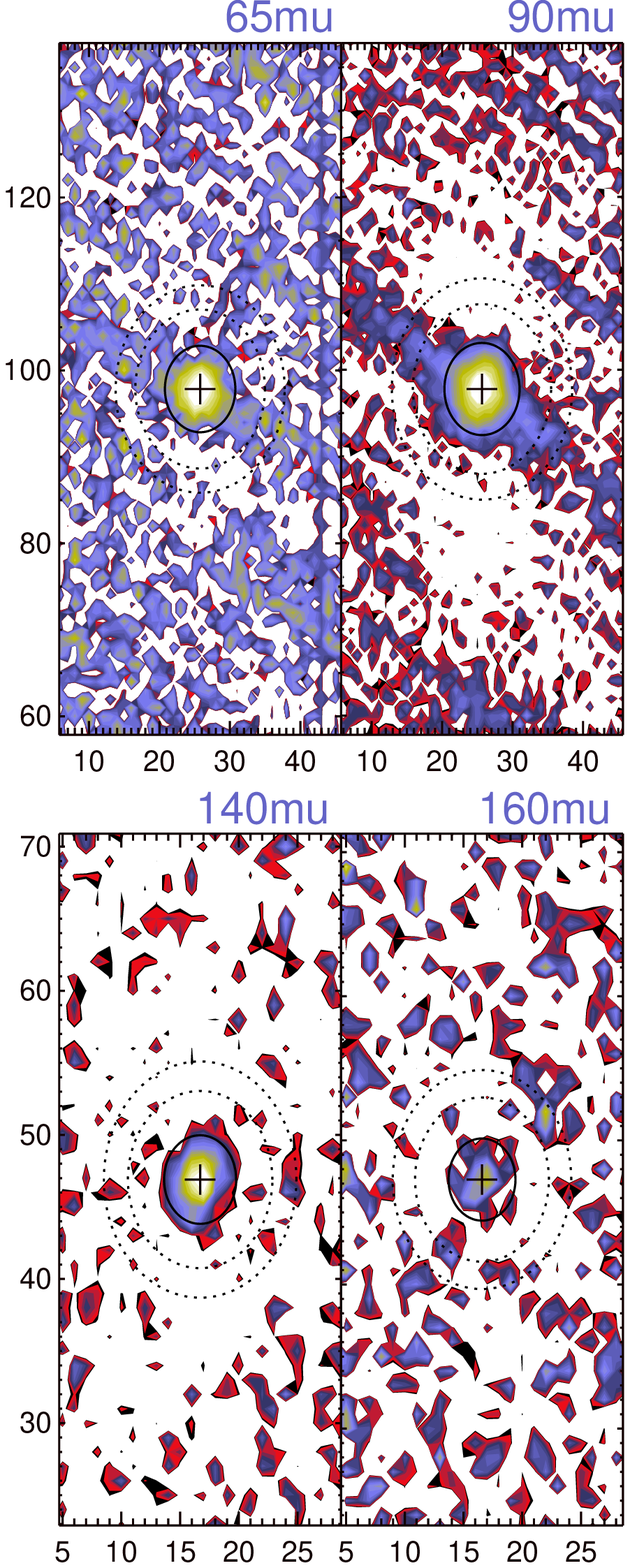}
\caption{(continued). For IRAS\,17395-0841 and IRAS\,19500-1709.}
\end{figure*}
\clearpage
\addtocounter{figure}{-1}

\begin{figure*}
\centering
\includegraphics[bb=30 140 300 800,width=5cm,height=8.6cm,clip]{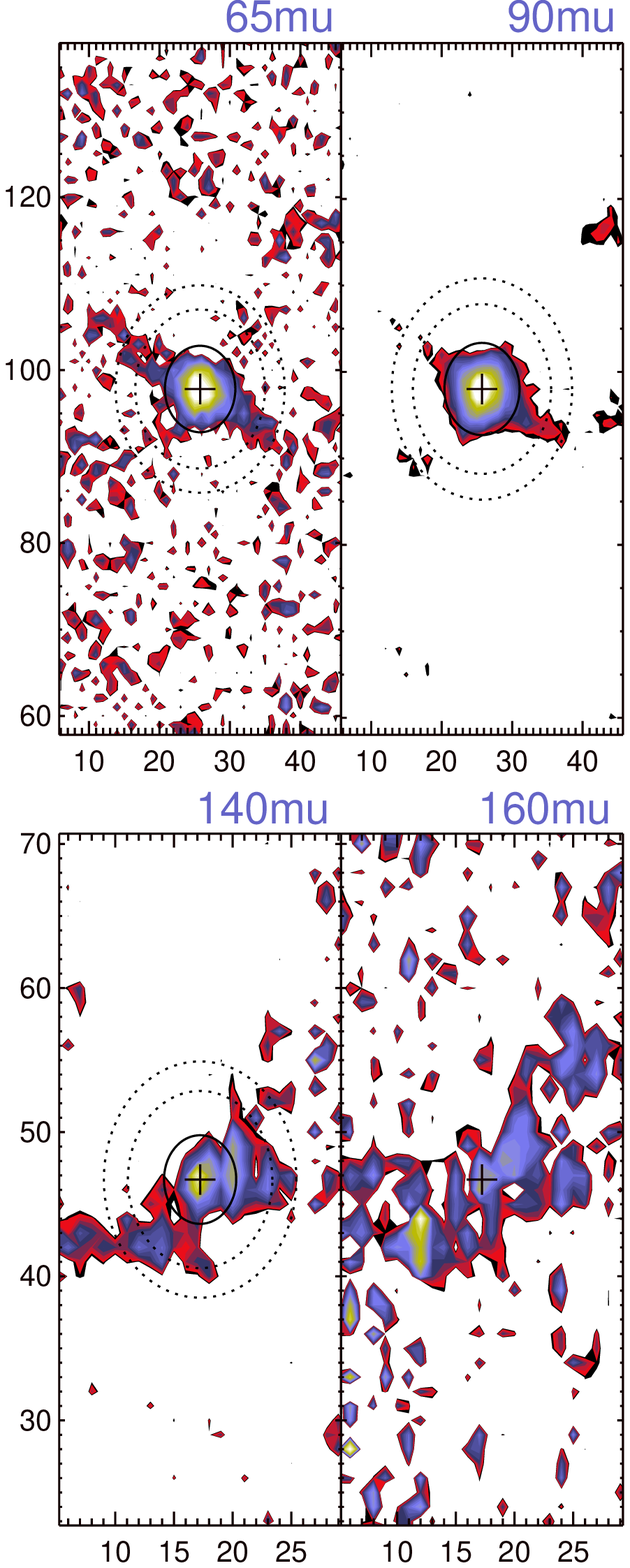}
\includegraphics[bb=30 140 300 800,width=5cm,height=8.6cm,clip]{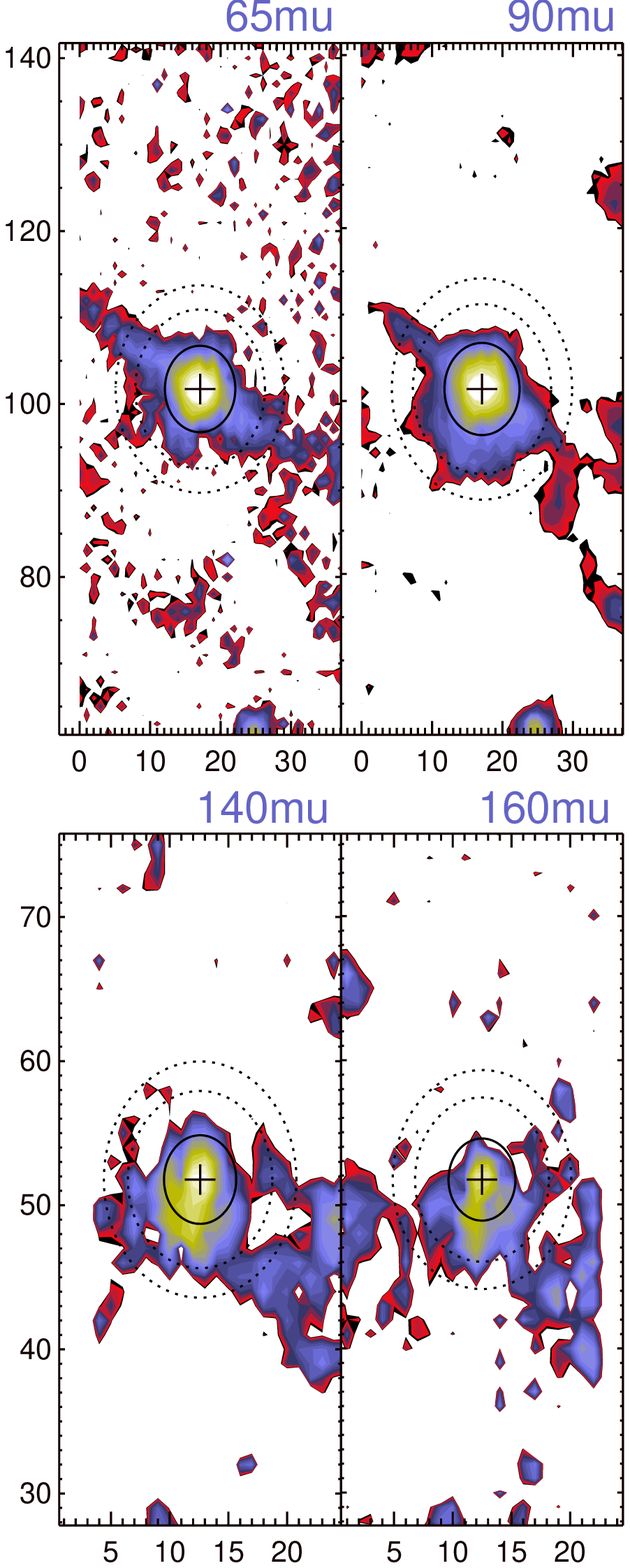}
\caption{%
(continued). For  IRAS\,20104+1950 and IRAS\,21046+4739.
The long wavelength FIS maps of IRAS\,21046+4739 show extended emission - small arc - just below the central source. The extend is a few arcminutes.
It is tentatively visible in the 90~$\mu$m image.}
\end{figure*}
\addtocounter{figure}{-1}

\begin{figure*}
\centering
\includegraphics[bb=30 140 300 800,width=5cm,height=8.6cm,clip]{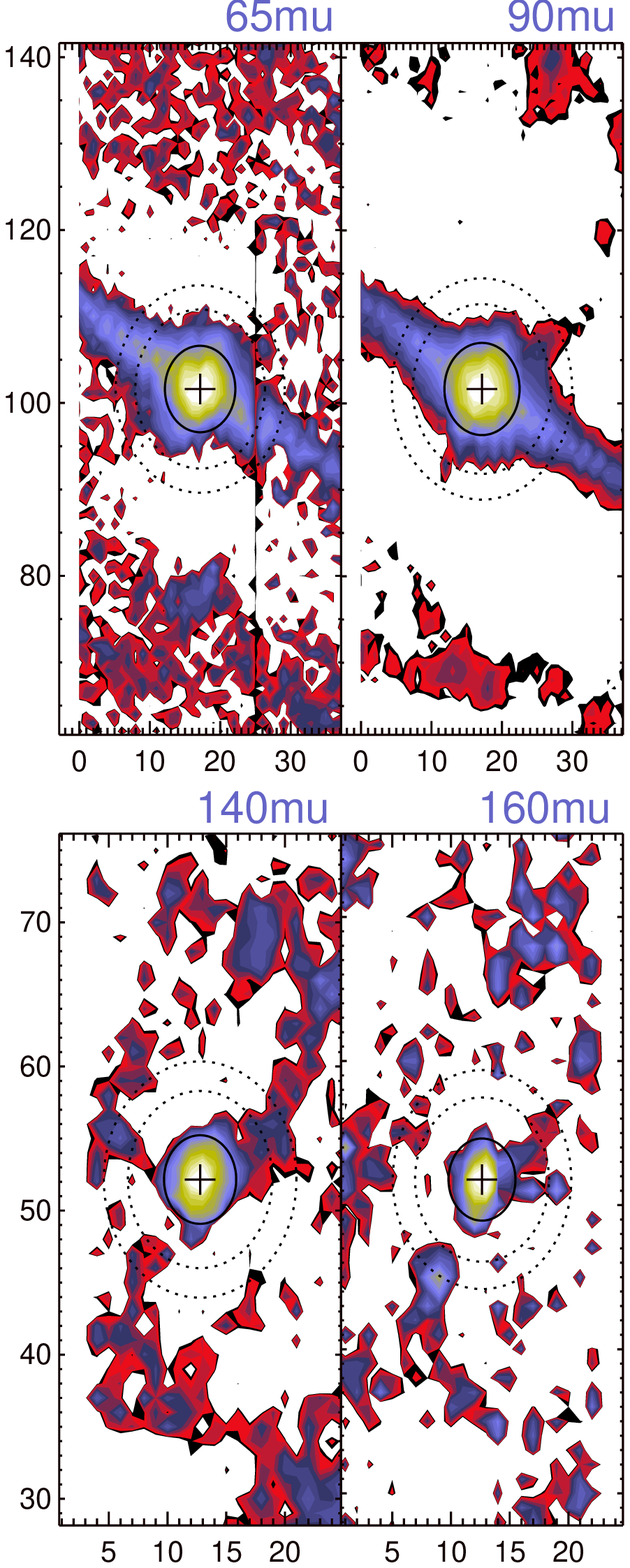}
\caption{%
(continued). For IRAS\,22036+5306.}
\label{fig:fismap13}
\end{figure*}

\clearpage
\section{Discussion}\label{sec:discussion}%

Individual modeling of the spectral energy distribution of these targets with a complex radiative transfer code (like DUSTY) 
is beyond the scope of this paper, which is focussed on the cold extended/detached circumstellar dust emission,
as the new photometry points (of the central source) do not add significant constraints 
to the problem with respect to previous studies.
Furthermore, the comparison of our AKARI FIS maps with simulated images (or SEDs) of the expected extended dust
emission seems to be inappropiate due to the lack of a clear detection of extended
emission in the AKARI images; such models (\eg, 2-DUST; \citealt{2003ApJ...586.1338U}) also have
 their own limitations and assumptions, as well as many free parameters.
In particular, the uncertain distances to the post-AGB objects in our sample 
affect the model results. Thus, we adopt here a simple dust model, together with the upper limits derived from 
the new FIS maps (Table~\ref{tb:photometry}), in order to give first order estimates on upper limits on the detected dust (and gas) mass
in the extended envelopes of these evolved objects.

\subsection{Upper limits on the detected dust+gas mass in extended envelopes}

Interstellar and circumstellar dust emits primarily in the infrared. Assuming that the observed flux
comes from an optically thin emission region the total mass of the dust and the associated gas can be estimated, with a few
assumptions about the basic properties of the dust, from the following equation (see \eg\ \citealt{Li2005}, Ueta et al. 2010):
\begin{equation}
M_{\rm gas+dust}\ [g] 
  \approx\ \frac{d^2\ F_\nu\ \lambda^2}{2\ k\ T_{\rm dust}\ \kappa_{\rm abs}(\lambda)}\ g2d \label{eq:dustmass}
\end{equation}
with the flux $F_\nu$ in erg cm$^{-2}$ s$^{-1}$ Hz$^{-1}$ ($F_\nu = 10^{-23} F$(Jy)), the distance $d$ in cm, 
the wavelength $\lambda$ in cm, the dust temperature $T_{\rm dust}$ in~K, the specific opacity or dust mass absorption
coefficient, \emph{i.e.} opacity, $\kappa_{\rm abs}(\lambda)$ in cm$^2$ g$^{-1}$
and the Boltzman constant in erg K$^{-1}$. 
For the gas-to-dust mass ratio, $g2d$ we take 160 for oxygen-rich stars and 400 for carbon-rich stars 
(see \eg, \citealt{1985ApJ...293..273K}, \citealt{HerasHony2005}).
Apart from the uncertainties in the distance large uncertainties result from poor knowledge of grain properties
as expressed through $\kappa_{\rm abs}$ (\citealt{Li2005}).
$F_\nu$ is the $3\sigma$-upper limit on the flux, at 90 or 140~$\mu$m, detectable in a shell of 3~pc diameter around the central star (\citealt{2002ApJ...571..880V}).
We assume a homogenous distribution across an annulus with an inner radius of 1\arcmin\ and outer radius of ($5157/d$)\arcmin, respectively.
The flux noise levels measured from the maps are given in Table~\ref{tb:photometry}.
For the distance we take values either from the literature or, where this is unavailable, we assume $d =1$--$2$~kpc (Table~\ref{tb:dustmass}). 

Values for opacity $\kappa$ are strongly dependent on (the composition of) the dust model or laboratory analog and can differ by an order of magnitude between studies.
For C-rich post-AGB stars with PAH features that have a high degree of hydrogenation a common choice is either
BE-type HAC dust from \citet{1995A&AS..113..561C} or the Silicate-Graphite-PAH dust model from \citet{2001ApJ...554..778L}.
The former give $\kappa(90\mu {\rm m})$ = 976~cm$^2$~g$^{-1}$, while the latter gives,
via $\kappa_{\rm abs} = 2.92\ 10^5\cdot \lambda^{-2}$, $\kappa(90\mu {\rm m})$ = 36~cm$^2$~g$^{-1}$.
Conservative dust mass limits are obtained by adopting the latter values for $\kappa$. These are also taken if we do not know the chemistry of the envelope.
For the sources with an oxygen-rich dust or mixed chemistry in the envelope 
we adopt $\kappa_{\rm abs}$ for astronomical silicates (\ie\ 63~cm$^2$~g$^{-1}$ at 90~$\mu$m; Draine \& Li 1984).

In computing the total extended dust and gas mass we consider relatively cold dust grains with $T_{\rm dust}$ = 20~K as these correspond
to old mass-loss events which has already cooled down and drifted farthest away (\emph{e.g.} \citealt{2006ApJ...640..829G}).
The resulting upper limits to the `hidden' dust \& gas mass in the extended shells around the observed stars are reported in Table~\ref{tb:dustmass}.
From equation~\ref{eq:dustmass} we remark that the derived upper limit on the envelope mass reduces proportionaly to an increase in dust grain temperature
and/or the dust opacity. The dust mass limits are proportional to the adapted gas-to-dust ratio and the distance squared.
In particular the latter can introduce significant uncertainties.

\begin{table*}[t!]
\caption{Derived dust+gas mass upper limits at 90~$\mu$m.}
\label{tb:dustmass}
\centering
\begin{tabular}{lllr}\tableline\tableline
                &  $d$         	& Chemistry/Dust &  Dust+gas mass upper limit\\	 									  
IRAS name	&  (kpc)	& 		 &   ($M_\odot$)       \\ \tableline			       
07134$+$1005    &  2.4$^{(1)}$	&   C		 &    1.1     	       \\ 
09425$-$6040    &  1 -- 2	&   O      	 &    0.2  -- 0.88     \\				      
10178$-$5958    &  1 -- 2	&   -		 &    1.9  -- 7.5      \\ 
10197$-$5750    &  1 -- 2	&   O    	 &    3.0  -- 12.0     \\ 
11478$-$5654    &  0.6$^{(2)}$	&   C		 &    0.39    	       \\				      
13428$-$6232    &  1 -- 2	&   -		 &   13.9  -- 55.7     \\				      
15318$-$7144    &  0.9$^{(2)}$	&   O		 &    0.31    	       \\				      
17119$-$5926    &  1 -- 2	&   -		 &    0.16 -- 0.64     \\				      
17395$-$0841    &  1.0$^{(3)}$	&   O		 &    0.08    	       \\ 
19500$-$1709    &  1 -- 2	&   C		 &    1.3  -- 5.4      \\				      
20104$+$1950    &  1.8$^{(2)}$	&   C		 &    0.21    	       \\				      
21046$+$4739    &  1.3$^{(2)}$	&   C		 &    0.53    	       \\				      
22036$+$5306    &  2$^{(4)}$	&   O		 &    0.12             \\				      
\tableline																		  
\end{tabular}
\begin{tabular}{l}
References: (1) {\citealt{2003A&A...402..211H}}; (2) \citealt{1989ApJ...345..306L}; (3) {\citealt{2003A&A...404..305G}}; 
(4) {\citealt{2008Ap&SS.313..241S}}
\end{tabular}
\tablenotetext{}{Flux limits are obtained from the observed FIS detection limits and the angular size
of the envelope assuming a radius of 3.0~pc (approximately between 5 and 10\arcmin) and excluding the inner 1\arcmin. 
See text for for details on the calculations and the adopted dust properties. 
In particular, we note the dependence of the dust mass with the the distance squared. 
We adopt gas-to-dust ratios of 160 for oxygen-rich dust and 400 for carbon-rich dust (see text). 
To obtain mass limits for a more generic gas-to-dust ratio of 200 divide the carbon-dust limits by two.
}
\end{table*}

The 90 $\mu$m FIS scan maps together with the aforementioned adopted grain properties provide upper limits to the total 
dust and gas mass between 0.1 and 7.5~$M_\odot$ (excluding IRAS\,10197-5750 and IRAS\,13428-6232 which show high levels of background noise). 
Noting the limits of the dust model (see above), we find that the dust and gas mass limits compare well with the total mass
($\sim$1 to 6~$M_\odot$) predicted to be lost by a star, with a certain initial mass, at the end of the AGB phase, as predicted by \citet{2002ApJ...571..880V}.
The estimated dust and gas mass limits are generally high, indicating that the envelope masses of these evolved 
stars should have been large in order to have been clearly detected with the FIS.
Nonetheless, for several cases the extended emission can be constrained to less than 1~$M_\odot$.

\subsection{Emission features in the FIS maps}
We briefly highlight several tantalizing, although tentative, features that can be discerned in the AKARI/FIS scan maps of several individual objects.
The \LW\ maps of IRAS\,09425-6040 show an arc-like dust emission  structure 10\arcmin\ upwards (in the direction of the ecliptic) of the central source. 
It is also visible, though very faint, in the \SW\ image. This structure is brightest at 140~$\mu$m,  indicating  a dust temperature of $\sim$20~K.
The 140 and 160~$\mu$m maps IRAS\,10178-5958 show some extended emission in two arcs 4\arcmin\ downwards and 8\arcmin\ upwards of the central source.
There is some indication of this emission at 90~$\mu$m but not at 65~$\mu$m, indicating that this is cooler dust, with a temperature of less than 20~K.
The 140~$\mu$m flux toward IRAS\,10197-5750 may be underestimated due to extended `cirrus' emission contaminating
the background measurements. The structure in the two SW bands is very similar (two arcs at 10\arcmin\ distance from the central source) while
very different from the two \LW\ bands in which a dust bar perpendicular to the direction of the ecliptic is observed. 
Thus, there appear to be two distinct structures that exhibit different average dust temperatures.
The FIS maps of IRAS\,13428-6232 show a wealth of structure in the far-infrared emission that show a higher flux at longer
wavelengths indicating a lower temperature of this diffuse `cirrus' dust. The central source is only detected in the SW bands.
The \LW\ bands show an indication of a central point source but the extended structure makes it all but impossible to do photometry.
Whether these structures are detached shells at about 7\arcmin\ or related to diffuse Galactic dust emission can not be discerned with these observations.
The FIS maps of IRAS\,21046+4739 show some indication of extended - cold - emission at the 'south-east' of the central source.
This asymmetric enhanced emission could be a sign of a shock interface between (recently) expelled material and the local ISM material.
IRAS\,21046+4739 appears to be the only target in which the extended structure also causes a faint
far-infrared emission excess in the \LW\ radial profile at radii between 30 and 130\arcsec.

\subsection{Disrepant far-infrared point source flux densities}

The FIR flux densities of IRAS\,17119-5926 and IRAS\,17395-0841 are significantly lower (by a factor two) than predicted from the
IRAS data. The FIS images show no obvious source of contamination in the \SW\ bands just as IRAS does not give a high value for cirrus emission. 
IRAS\,17119-5926, however, is a peculiar source that is thought to have evolved only recently, in the last few decades, from a post-AGB into a PN \citep{1993A&A...267L..19P}. 
Therefore, variability in its infrared flux might occur due to recent mass loss event(s) or evolution of the central star.
This is supported also by the lower continuum level from the SWS spectrum. 
The latter appearing in fact more consistent with the new FIS flux densities.
For IRAS\,17395-0841 the emission in the \LW\ images does not appear to be directly related to the point source.
The IRAS point source catalog (PSC) lists a relatively high value for the cirrus emission at 100$\mu$m.
Four other sources in our sample have high cirrus values in the IRAS PSC but their FIS fluxes match well with those given by IRAS.
Also, from the SEDs we find that nebular emission, as expected for post-AGB/proto-PNe, contributes very little to the observed broad band flux.

\section{Concluding remarks}\label{sec:conclusion}%

Our observations provide imaging of post-AGB and (P)PNe beyond 100$\mu$m for which only limited information is currently available. 
The AKARI FIS photometry is generally consistent, considering also color correction, 
differences in filter curves and absolute calibration uncertainties, with the available SEDs, based in the infrared - primarily on IRAS and ISO data.
We also conclude that the far-infrared point source emission detected toward our sample of post-AGB stars and (P)PNe
originates from a compact region unresolved by the FIS.

The observed background emission (measured for each map) was used to derive first order upper limits on the mass in `cold' extended emission. 
The derived dust and gas masses are strongly dependent on, in particular, the adopted distance, and adopted grain size and emissivity 
(both of which affect the equilibrium dust temperature).
For a few objects we note potentially diffuse infrared emission structures that could be related to detached shells or wind-ISM interaction;
IRAS\,10178-5958 shows two arcs at 4 and 8\arcmin\ from the central object at 140 and 160~$\mu$m, IRAS\,10197-5750 shows two shells at
10\arcmin\ in the short wavelength filters and a peculiar cold dust bar at the longest wavelengths,
and IRAS\,21046+4739 reveals both in the FIS images and the radial profiles extended emission near the central source ($<$3\arcmin) at
140 and 160~$\mu$m.

Further studies of the extended emission (in particular for the objects with tentative detections of mass-loss events) will 
require multi-wavelength observations to obtain not only an estimate of the total dust and gas mass but also some insight into the basic properties 
(temperature, size, composition) of grains in these extended shells.
Imaging with both improved sensitivity and higher spatial resolution are required to observe extended structures due to cold 
dust associated with the sources presented above. In particular the capabilities of the Herschel Space Observatory are favorable to 
detecting the coldest dust components in extended areas and at the high spatial resolution required to disentangle circumstellar dust from 
cirrus dust.

\acknowledgments
The AKARI FIS team (M. Shirahata, S. Matsuura, S. Makiuti, I. Yamamura, and E. Verdugo)
is warmly thanked for providing helpful information on the data processing.
This work is based on observations with AKARI, a JAXA project with the participation of ESA.
D.A.G.H. and A.M.  acknowledge support for this work provided by the Spanish Ministry
of Science and Innovation (MICINN) under the 2008 Juan de la Cierva Programme
and under grant AYA-2007-64748. 
N.L.J.C. thanks S.\,Hony for fruitful discussions.
We thank the referee for helping to improve this paper.
We thank R.\.Szczerba for making the {\it Toru\'n catalogue of Galactic post-AGB and related objects} available to the community.
This research made use of the IRAS Point Source Catalog (version 2.1).
This research has made use of the SIMBAD database, operated at CDS, Strasbourg, France.

{\it Facilities:} \facility{AKARI (FIS)}

\end{document}